\documentclass[conference,letterpaper]{IEEEtran}
\usepackage[letterpaper, left=0.680in, right=0.647in, bottom=1in, top=0.71in]{geometry}

\IEEEoverridecommandlockouts\IEEEoverridecommandlockouts

\usepackage{cite}
\usepackage{amsmath,amssymb,amsfonts}
\usepackage{graphicx}
\usepackage{textcomp}
\usepackage{xcolor}
\usepackage{algorithm}
\usepackage{algorithmic}
\usepackage{balance}
\usepackage{dsfont}

\usepackage{setspace}
\usepackage{amsthm}
\theoremstyle{definition}
\newtheorem{mydef}{Definition}

\usepackage{pgfplots}
\pgfplotsset{width=8cm,compat=1.9}

\usepgfplotslibrary{external}
\DeclareMathOperator*{\argmin}{arg\,min}

\usepackage{amssymb}
\usepackage{amsfonts}
\usepackage{mathrsfs}
\usepackage{xspace}
\usepackage{bm}
\usepackage{upgreek}

\newcommand{\safemath}[2]{\newcommand{#1}{\ensuremath{#2}\xspace}}

\safemath{\bma}{\mathbf{a}}
\safemath{\bmb}{\mathbf{b}}
\safemath{\bmc}{\mathbf{c}}
\safemath{\bmd}{\mathbf{d}}
\safemath{\bme}{\mathbf{e}}
\safemath{\bmf}{\mathbf{f}}
\safemath{\bmg}{\mathbf{g}}
\safemath{\bmh}{\mathbf{h}}
\safemath{\bmi}{\mathbf{i}}
\safemath{\bmj}{\mathbf{j}}
\safemath{\bmk}{\mathbf{k}}
\safemath{\bml}{\mathbf{l}}
\safemath{\bmm}{\mathbf{m}}
\safemath{\bmn}{\mathbf{n}}
\safemath{\bmo}{\mathbf{o}}
\safemath{\bmp}{\mathbf{p}}
\safemath{\bmq}{\mathbf{q}}
\safemath{\bmr}{\mathbf{r}}
\safemath{\bms}{\mathbf{s}}
\safemath{\bmt}{\mathbf{t}}
\safemath{\bmu}{\mathbf{u}}
\safemath{\bmv}{\mathbf{v}}
\safemath{\bmw}{\mathbf{w}}
\safemath{\bmx}{\mathbf{x}}
\safemath{\bmy}{\mathbf{y}}
\safemath{\bmz}{\mathbf{z}}
\safemath{\bmzero}{\mathbf{0}}
\safemath{\bmone}{\mathbf{1}}

\bmdefine{\biad}{a}
\bmdefine{\bibd}{b}
\bmdefine{\bicd}{c}
\bmdefine{\bidd}{d}
\bmdefine{\bied}{e}
\bmdefine{\bifd}{f}
\bmdefine{\bigd}{g}
\bmdefine{\bihd}{h}
\bmdefine{\biid}{i}
\bmdefine{\bijd}{j}
\bmdefine{\bikd}{k}
\bmdefine{\bild}{l}
\bmdefine{\bimd}{m}
\bmdefine{\bind}{n}
\bmdefine{\biod}{o}
\bmdefine{\bipd}{p}
\bmdefine{\biqd}{q}
\bmdefine{\bird}{r}
\bmdefine{\bisd}{s}
\bmdefine{\bitd}{t}
\bmdefine{\biud}{u}
\bmdefine{\bivd}{v}
\bmdefine{\biwd}{w}
\bmdefine{\bixd}{x}
\bmdefine{\biyd}{y}
\bmdefine{\bizd}{z}

\bmdefine{\bixid}{\xi}
\bmdefine{\bilambdad}{\lambda}
\bmdefine{\bimud}{\mu}
\bmdefine{\bithetad}{\theta}
\bmdefine{\biphid}{\phi}
\bmdefine{\bideltad}{\delta}

\safemath{\bmia}{\biad}
\safemath{\bmib}{\bibd}
\safemath{\bmic}{\bicd}
\safemath{\bmid}{\bidd}
\safemath{\bmie}{\bied}
\safemath{\bmif}{\bifd}
\safemath{\bmig}{\bigd}
\safemath{\bmih}{\bihd}
\safemath{\bmii}{\biid}
\safemath{\bmij}{\bijd}
\safemath{\bmik}{\bikd}
\safemath{\bmil}{\bild}
\safemath{\bmim}{\bimd}
\safemath{\bmin}{\bind}
\safemath{\bmio}{\biod}
\safemath{\bmip}{\bipd}
\safemath{\bmiq}{\biqd}
\safemath{\bmir}{\bird}
\safemath{\bmis}{\bisd}
\safemath{\bmit}{\bitd}
\safemath{\bmiu}{\biud}
\safemath{\bmiv}{\bivd}
\safemath{\bmiw}{\biwd}
\safemath{\bmix}{\bixd}
\safemath{\bmiy}{\biyd}
\safemath{\bmiz}{\bizd}

\safemath{\bmxi}{\bixid}
\safemath{\bmlambda}{\bilambdad}
\safemath{\bmmu}{\bimud}
\safemath{\bmtheta}{\bithetad}
\safemath{\bmphi}{\biphid}
\safemath{\bmdelta}{\bideltad}

\safemath{\bA}{\mathbf{A}}
\safemath{\bB}{\mathbf{B}}
\safemath{\bC}{\mathbf{C}}
\safemath{\bD}{\mathbf{D}}
\safemath{\bE}{\mathbf{E}}
\safemath{\bF}{\mathbf{F}}
\safemath{\bG}{\mathbf{G}}
\safemath{\bH}{\mathbf{H}}
\safemath{\bI}{\mathbf{I}}
\safemath{\bJ}{\mathbf{J}}
\safemath{\bK}{\mathbf{K}}
\safemath{\bL}{\mathbf{L}}
\safemath{\bM}{\mathbf{M}}
\safemath{\bN}{\mathbf{N}}
\safemath{\bO}{\mathbf{O}}
\safemath{\bP}{\mathbf{P}}
\safemath{\bQ}{\mathbf{Q}}
\safemath{\bR}{\mathbf{R}}
\safemath{\bS}{\mathbf{S}}
\safemath{\bT}{\mathbf{T}}
\safemath{\bU}{\mathbf{U}}
\safemath{\bV}{\mathbf{V}}
\safemath{\bW}{\mathbf{W}}
\safemath{\bX}{\mathbf{X}}
\safemath{\bY}{\mathbf{Y}}
\safemath{\bZ}{\mathbf{Z}}

\safemath{\bZero}{\mathbf{0}}
\safemath{\bOne}{\mathbf{1}}
\safemath{\bDelta}{\mathbf{\Delta}}
\safemath{\bLambda}{\mathbf{\UpLambda}}
\safemath{\bPhi}{\mathbf{\Upphi}}
\safemath{\bSigma}{\mathbf{\Upsigma}}
\safemath{\bOmega}{\mathbf{\Upomega}}
\safemath{\bTheta}{\mathbf{\Uptheta}}

\bmdefine{\biAd}{A}
\bmdefine{\biBd}{B}
\bmdefine{\biCd}{C}
\bmdefine{\biDd}{D}
\bmdefine{\biEd}{E}
\bmdefine{\biFd}{F}
\bmdefine{\biGd}{G}
\bmdefine{\biHd}{H}
\bmdefine{\biId}{I}
\bmdefine{\biJd}{J}
\bmdefine{\biKd}{K}
\bmdefine{\biLd}{L}
\bmdefine{\biMd}{M}
\bmdefine{\biOd}{N}
\bmdefine{\biPd}{O}
\bmdefine{\biQd}{P}
\bmdefine{\biRd}{R}
\bmdefine{\biSd}{S}
\bmdefine{\biTd}{T}
\bmdefine{\biUd}{U}
\bmdefine{\biVd}{V}
\bmdefine{\biWd}{W}
\bmdefine{\biXd}{X}
\bmdefine{\biYd}{Y}
\bmdefine{\biZd}{Z}

\bmdefine{\biDelta}{\Delta}
\bmdefine{\biLambda}{\Lambda}
\bmdefine{\biPhi}{\Phi}
\bmdefine{\biSigma}{\Sigma}
\bmdefine{\biOmega}{\Omega}
\bmdefine{\biTheta}{\Theta}

\safemath{\bimA}{\biAd}
\safemath{\bimB}{\biBd}
\safemath{\bimC}{\biCd}
\safemath{\bimD}{\biDd}
\safemath{\bimE}{\biEd}
\safemath{\bimF}{\biFd}
\safemath{\bimG}{\biGd}
\safemath{\bimH}{\biHd}
\safemath{\bimI}{\biId}
\safemath{\bimJ}{\biJd}
\safemath{\bimK}{\biKd}
\safemath{\bimL}{\biLd}
\safemath{\bimM}{\biMd}
\safemath{\bimN}{\biNd}
\safemath{\bimO}{\biOd}
\safemath{\bimP}{\biPd}
\safemath{\bimQ}{\biQd}
\safemath{\bimR}{\biRd}
\safemath{\bimS}{\biSd}
\safemath{\bimT}{\biTd}
\safemath{\bimU}{\biUd}
\safemath{\bimV}{\biVd}
\safemath{\bimW}{\biWd}
\safemath{\bimX}{\biXd}
\safemath{\bimY}{\biYd}
\safemath{\bimZ}{\biZd}

\safemath{\bimDelta}{\biDelta}
\safemath{\bimLambda}{\biLambda}
\safemath{\bimPhi}{\biPhi}
\safemath{\bimSigma}{\biSigma}
\safemath{\bimOmega}{\biOmega}
\safemath{\bimTheta}{\biTheta}

\safemath{\setA}{\mathcal{A}}
\safemath{\setB}{\mathcal{B}}
\safemath{\setC}{\mathcal{C}}
\safemath{\setD}{\mathcal{D}}
\safemath{\setE}{\mathcal{E}}
\safemath{\setF}{\mathcal{F}}
\safemath{\setG}{\mathcal{G}}
\safemath{\setH}{\mathcal{H}}
\safemath{\setI}{\mathcal{I}}
\safemath{\setJ}{\mathcal{J}}
\safemath{\setK}{\mathcal{K}}
\safemath{\setL}{\mathcal{L}}
\safemath{\setM}{\mathcal{M}}
\safemath{\setN}{\mathcal{N}}
\safemath{\setO}{\mathcal{O}}
\safemath{\setP}{\mathcal{P}}
\safemath{\setQ}{\mathcal{Q}}
\safemath{\setR}{\mathcal{R}}
\safemath{\setS}{\mathcal{S}}
\safemath{\setT}{\mathcal{T}}
\safemath{\setU}{\mathcal{U}}
\safemath{\setV}{\mathcal{V}}
\safemath{\setW}{\mathcal{W}}
\safemath{\setX}{\mathcal{X}}
\safemath{\setY}{\mathcal{Y}}
\safemath{\setZ}{\mathcal{Z}}
\safemath{\emptySet}{\varnothing}

\safemath{\colA}{\mathscr{A}}
\safemath{\colB}{\mathscr{B}}
\safemath{\colC}{\mathscr{C}}
\safemath{\colD}{\mathscr{D}}
\safemath{\colE}{\mathscr{E}}
\safemath{\colF}{\mathscr{F}}
\safemath{\colG}{\mathscr{G}}
\safemath{\colH}{\mathscr{H}}
\safemath{\colI}{\mathscr{I}}
\safemath{\colJ}{\mathscr{J}}
\safemath{\colK}{\mathscr{K}}
\safemath{\colL}{\mathscr{L}}
\safemath{\colM}{\mathscr{M}}
\safemath{\colN}{\mathscr{N}}
\safemath{\colO}{\mathscr{O}}
\safemath{\colP}{\mathscr{P}}
\safemath{\colQ}{\mathscr{Q}}
\safemath{\colR}{\mathscr{R}}
\safemath{\colS}{\mathscr{S}}
\safemath{\colT}{\mathscr{T}}
\safemath{\colU}{\mathscr{U}}
\safemath{\colV}{\mathscr{V}}
\safemath{\colW}{\mathscr{W}}
\safemath{\colX}{\mathscr{X}}
\safemath{\colY}{\mathscr{Y}}
\safemath{\colZ}{\mathscr{Z}}

\safemath{\opA}{\mathbb{A}}
\safemath{\opB}{\mathbb{B}}
\safemath{\opC}{\mathbb{C}}
\safemath{\opD}{\mathbb{D}}
\safemath{\opE}{\mathbb{E}}
\safemath{\opF}{\mathbb{F}}
\safemath{\opG}{\mathbb{G}}
\safemath{\opH}{\mathbb{H}}
\safemath{\opI}{\mathbb{I}}
\safemath{\opJ}{\mathbb{J}}
\safemath{\opK}{\mathbb{K}}
\safemath{\opL}{\mathbb{L}}
\safemath{\opM}{\mathbb{M}}
\safemath{\opN}{\mathbb{N}}
\safemath{\opO}{\mathbb{O}}
\safemath{\opP}{\mathbb{P}}
\safemath{\opQ}{\mathbb{Q}}
\safemath{\opR}{\mathbb{R}}
\safemath{\opS}{\mathbb{S}}
\safemath{\opT}{\mathbb{T}}
\safemath{\opU}{\mathbb{U}}
\safemath{\opV}{\mathbb{V}}
\safemath{\opW}{\mathbb{W}}
\safemath{\opX}{\mathbb{X}}
\safemath{\opY}{\mathbb{Y}}
\safemath{\opZ}{\mathbb{Z}}
\safemath{\opZero}{\mathbb{O}}
\safemath{\identityop}{\opI}

\safemath{\veca}{\bma}
\safemath{\vecb}{\bmb}
\safemath{\vecc}{\bmc}
\safemath{\vecd}{\bmd}
\safemath{\vece}{\bme}
\safemath{\vecf}{\bmf}
\safemath{\vecg}{\bmg}
\safemath{\vech}{\bmh}
\safemath{\veci}{\bmi}
\safemath{\vecj}{\bmj}
\safemath{\veck}{\bmk}
\safemath{\vecl}{\bml}
\safemath{\vecm}{\bmm}
\safemath{\vecn}{\bmn}
\safemath{\veco}{\bmo}
\safemath{\vecp}{\bmp}
\safemath{\vecq}{\bmq}
\safemath{\vecr}{\bmr}
\safemath{\vecs}{\bms}
\safemath{\vect}{\bmt}
\safemath{\vecu}{\bmu}
\safemath{\vecv}{\bmv}
\safemath{\vecw}{\bmw}
\safemath{\vecx}{\bmx}
\safemath{\vecy}{\bmy}
\safemath{\vecz}{\bmz}

\safemath{\veczero}{\bmzero}
\safemath{\vecone}{\bmone}
\safemath{\vecxi}{\bmxi}
\safemath{\veclambda}{\bmlambda}
\safemath{\vecmu}{\bmmu}
\safemath{\vectheta}{\bmtheta}
\safemath{\vecphi}{\bmphi}
\safemath{\vecdelta}{\bmdelta}

\safemath{\matA}{\bA}
\safemath{\matB}{\bB}
\safemath{\matC}{\bC}
\safemath{\matD}{\bD}
\safemath{\matE}{\bE}
\safemath{\matF}{\bF}
\safemath{\matG}{\bG}
\safemath{\matH}{\bH}
\safemath{\matI}{\bI}
\safemath{\matJ}{\bJ}
\safemath{\matK}{\bK}
\safemath{\matL}{\bL}
\safemath{\matM}{\bM}
\safemath{\matN}{\bN}
\safemath{\matO}{\bO}
\safemath{\matP}{\bP}
\safemath{\matQ}{\bQ}
\safemath{\matR}{\bR}
\safemath{\matS}{\bS}
\safemath{\matT}{\bT}
\safemath{\matU}{\bU}
\safemath{\matV}{\bV}
\safemath{\matW}{\bW}
\safemath{\matX}{\bX}
\safemath{\matY}{\bY}
\safemath{\matZ}{\bZ}
\safemath{\matzero}{\bmzero}

\safemath{\matDelta}{\bDelta}
\safemath{\matLambda}{\bLambda}
\safemath{\matPhi}{\bPhi}
\safemath{\matSigma}{\bSigma}
\safemath{\matOmega}{\bOmega}
\safemath{\matTheta}{\bTheta}

\safemath{\matidentity}{\matI}
\safemath{\matone}{\matO}

\safemath{\rnda}{A}
\safemath{\rndb}{B}
\safemath{\rndc}{C}
\safemath{\rndd}{D}
\safemath{\rnde}{E}
\safemath{\rndf}{F}
\safemath{\rndg}{G}
\safemath{\rndh}{H}
\safemath{\rndi}{I}
\safemath{\rndj}{J}
\safemath{\rndk}{K}
\safemath{\rndl}{L}
\safemath{\rndm}{M}
\safemath{\rndn}{N}
\safemath{\rndo}{O}
\safemath{\rndp}{P}
\safemath{\rndq}{Q}
\safemath{\rndr}{R}
\safemath{\rnds}{S}
\safemath{\rndt}{T}
\safemath{\rndu}{U}
\safemath{\rndv}{V}
\safemath{\rndw}{W}
\safemath{\rndx}{X}
\safemath{\rndy}{Y}
\safemath{\rndz}{Z}

\safemath{\rveca}{\bimA}
\safemath{\rvecb}{\bimB}
\safemath{\rvecc}{\bimC}
\safemath{\rvecd}{\bimD}
\safemath{\rvece}{\bimE}
\safemath{\rvecf}{\bimF}
\safemath{\rvecg}{\bimG}
\safemath{\rvech}{\bimH}
\safemath{\rveci}{\bimI}
\safemath{\rvecj}{\bimJ}
\safemath{\rveck}{\bimK}
\safemath{\rvecl}{\bimL}
\safemath{\rvecm}{\bimM}
\safemath{\rvecn}{\bimN}
\safemath{\rveco}{\bomO}
\safemath{\rvecp}{\bimP}
\safemath{\rvecq}{\bimQ}
\safemath{\rvecr}{\bimR}
\safemath{\rvecs}{\bimS}
\safemath{\rvect}{\bimT}
\safemath{\rvecu}{\bimU}
\safemath{\rvecv}{\bimV}
\safemath{\rvecw}{\bimW}
\safemath{\rvecx}{\bimX}
\safemath{\rvecy}{\bimY}
\safemath{\rvecz}{\bimZ}

\safemath{\rvecxi}{\bmxi}
\safemath{\rveclambda}{\bmlambda}
\safemath{\rvecmu}{\bmmu}
\safemath{\rvectheta}{\bmtheta}
\safemath{\rvecphi}{\bmphi}

\safemath{\rmatA}{\bimA}
\safemath{\rmatB}{\bimB}
\safemath{\rmatC}{\bimC}
\safemath{\rmatD}{\bimD}
\safemath{\rmatE}{\bimE}
\safemath{\rmatF}{\bimF}
\safemath{\rmatG}{\bimG}
\safemath{\rmatH}{\bimH}
\safemath{\rmatI}{\bimI}
\safemath{\rmatJ}{\bimJ}
\safemath{\rmatK}{\bimK}
\safemath{\rmatL}{\bimL}
\safemath{\rmatM}{\bimM}
\safemath{\rmatN}{\bimN}
\safemath{\rmatO}{\bimO}
\safemath{\rmatP}{\bimP}
\safemath{\rmatQ}{\bimQ}
\safemath{\rmatR}{\bimR}
\safemath{\rmatS}{\bimS}
\safemath{\rmatT}{\bimT}
\safemath{\rmatU}{\bimU}
\safemath{\rmatV}{\bimV}
\safemath{\rmatW}{\bimW}
\safemath{\rmatX}{\bimX}
\safemath{\rmatY}{\bimY}
\safemath{\rmatZ}{\bimZ}

\safemath{\rmatDelta}{\bimDelta}
\safemath{\rmatLambda}{\bimLambda}
\safemath{\rmatPhi}{\bimPhi}
\safemath{\rmatSigma}{\bimSigma}
\safemath{\rmatOmega}{\bimOmega}
\safemath{\rmatTheta}{\bimTheta}

\usepackage{amssymb}
\usepackage{amsfonts}
\usepackage{mathrsfs}
\usepackage{xspace}
\usepackage{bm}
\usepackage{fancyref}
\usepackage{textcomp}

\usepackage{multirow}
\usepackage{stmaryrd}

\newenvironment{textbmatrix}{	\setlength{\arraycolsep}{2.5pt}%
								\big[\begin{matrix}}{\end{matrix}\big]%
								\raisebox{0.08ex}{\vphantom{M}}}

\def\be{\begin{equation}}
\def\ee{\end{equation}}
\def\een{\nonumber \end{equation}}
\def\mat{\begin{bmatrix}}
\def\emat{\end{bmatrix}}
\def\btm{\begin{textbmatrix}}
\def\etm{\end{textbmatrix}}

\def\ba#1\ea{\begin{align}#1\end{align}}
\def\bas#1\eas{\begin{align*}#1\end{align*}}
\def\bs#1\es{\begin{split}#1\end{split}} 
\def\bg#1\eg{\begin{gather}#1\end{gather}}
\def\bml#1\eml{\begin{multline}#1\end{multline}}
\def\bi#1\ei{\begin{itemize}#1\end{itemize}}

\newcommand{\subtext}[1]{\text{\fontfamily{cmr}\fontshape{n}\fontseries{m}\selectfont{}#1}}
\newcommand{\lefto}{\mathopen{}\left}

\DeclareMathOperator{\Exop}{\opE}			
\DeclareMathOperator{\Varop}{\mathrm{Var}}
\DeclareMathOperator{\Covop}{\mathrm{Cov}}
\DeclareMathOperator{\rot}{\nabla\times}		

\newcommand{\vecnorm}[1]{\lefto\lVert#1\right\rVert}		

\safemath{\dirac}{\delta}					
\safemath{\krond}{\dirac}					

\safemath{\upto}{\uparrow}
\safemath{\downto}{\downarrow}
\safemath{\iu}{j}							
\safemath{\ev}{\lambda}						
\safemath{\hilseqspace}{l^{2}}				
\newcommand{\banachfunspace}[1]{\setL^{#1}}	
\safemath{\hilfunspace}{\banachfunspace{2}}	

\safemath{\SNR}{\textsf{SNR}} 				
\safemath{\PAR}{\textsf{PAR}} 				
\safemath{\No}{N_0}							
\safemath{\Es}{E_s}							
\safemath{\Eb}{E_b}							
\safemath{\EbNo}{\frac{\Eb}{\No}}
\safemath{\EsNo}{\frac{\Es}{\No}}

\DeclareMathOperator{\CHop}{\ensuremath{\opH}} 
\safemath{\tvir}{\rndh_{\CHop}}				
\safemath{\tvtf}{\rndl_{\CHop}}				
\safemath{\spf}{\rnds_{\CHop}}				
\safemath{\bff}{H_{\CHop}}					

\safemath{\ircf}{r_{h}}						
\safemath{\tftvcf}{r_{s}}					
\safemath{\tfcf}{r_{l}}						
\safemath{\bfcf}{r_{H}}						

\safemath{\tcorr}{c_h}						
\safemath{\scf}{c_{s}}						
\safemath{\tfcorr}{c_{l}}					
\safemath{\fcorr}{c_{H}}						

\safemath{\mi}{I}							
\safemath{\capacity}{C}						

\safemath{\normal}{\mathcal{N}}			
\safemath{\jpg}{\mathcal{CN}}			
\safemath{\mchain}{\leftrightarrow}		
\newcommand{\AIC}{\ensuremath{\text{AIC}}} 

\safemath{\dB}{\,\mathrm{dB}}
\safemath{\dBm}{\,\mathrm{dBm}}
\safemath{\Hz}{\,\mathrm{Hz}}
\safemath{\kHz}{\,\mathrm{kHz}}
\safemath{\MHz}{\,\mathrm{MHz}}
\safemath{\GHz}{\,\mathrm{GHz}}
\safemath{\s}{\,\mathrm{s}}
\safemath{\ms}{\,\mathrm{ms}}
\safemath{\mus}{\,\mathrm{\text{\textmu}s}}
\safemath{\ns}{\,\mathrm{ns}}
\safemath{\ps}{\,\mathrm{ps}}
\safemath{\meter}{\,\mathrm{m}}
\safemath{\mm}{\,\mathrm{mm}}
\safemath{\cm}{\,\mathrm{cm}}
\safemath{\m}{\,\mathrm{m}}
\safemath{\W}{\,\mathrm{W}}
\safemath{\mW}{\, \mathrm{mW}}
\safemath{\J}{\,\mathrm{J}}
\safemath{\K}{\,\mathrm{K}}
\safemath{\bit}{\,\mathrm{bit}}
\safemath{\nat}{\,\mathrm{nat}}

\safemath{\define}{\triangleq}			

\safemath{\equivalent}{\sim}
\safemath{\distas}{\sim}					
\safemath{\sdiff}{\Delta}				

\safemath{\reals}{\mathbb{R}}
\safemath{\positivereals}{\reals_{+}}
\safemath{\integers}{\mathbb{Z}}
\safemath{\posint}{\integers_{+}}
\safemath{\naturals}{\mathbb{N}}
\safemath{\posnaturals}{\naturals_{+}}
\safemath{\complexset}{\mathbb{C}}
\safemath{\rationals}{\mathbb{Q}}

\newcommand*{\fancyrefapplabelprefix}{app}		
\newcommand*{\fancyrefthmlabelprefix}{thm}		
\newcommand*{\fancyreflemlabelprefix}{lem}		
\newcommand*{\fancyrefcorlabelprefix}{cor}		
\newcommand*{\fancyrefdeflabelprefix}{def}		
\newcommand*{\fancyrefproplabelprefix}{prop}	
\newcommand*{\fancyrefobslabelprefix}{obs}		
\newcommand*{\fancyrefalglabelprefix}{alg}		
\newcommand*{\fancyrefasmlabelprefix}{asm}	    
\newcommand*{\fancyreftbllabelprefix}{tab}	 

\frefformat{vario}{\fancyrefseclabelprefix}{Section~#1}
\frefformat{vario}{\fancyrefthmlabelprefix}{Theorem~#1}
\frefformat{vario}{\fancyreflemlabelprefix}{Lemma~#1}
\frefformat{vario}{\fancyrefcorlabelprefix}{Corollary~#1}
\frefformat{vario}{\fancyrefdeflabelprefix}{Definition~#1}
\frefformat{vario}{\fancyrefobslabelprefix}{Observation~#1}
\frefformat{vario}{\fancyrefasmlabelprefix}{Assumption~#1}
\frefformat{vario}{\fancyreffiglabelprefix}{Fig.~#1}
\frefformat{vario}{\fancyrefapplabelprefix}{Appendix~#1} 
\frefformat{vario}{\fancyrefproplabelprefix}{Proposition~#1}
\frefformat{vario}{\fancyrefalglabelprefix}{Algorithm~#1}
\frefformat{vario}{\fancyreftbllabelprefix}{Table~#1}
\frefformat{vario}{\fancyrefeqlabelprefix}{(#1)}

\newcommand{\splitatcommas}[1]{%
  \begingroup
  \ifnum\mathcode`,="8000
  \else
    \begingroup\lccode`~=`, \lowercase{\endgroup
      \edef~{\mathchar\the\mathcode`, \penalty0 \noexpand\hspace{0pt plus 1em}}%
    }\mathcode`,="8000
  \fi
  #1%
  \endgroup
}

\usepackage{stackengine}
\stackMath
\newcommand\tsup[2][2]{%
 \def\useanchorwidth{T}%
  \ifnum#1>1%
    \stackon[-1.3ex]{\tsup[\numexpr#1-1\relax]{#2}}{\mathchar"307E}%
  \else%
    \stackon[-1ex]{#2}{\mathchar"307E}%
  \fi%
}

\ifCLASSINFOpdf
\else
\fi

\renewcommand{\baselinestretch}{1.049}
\begin{document}
%

\title{Privacy-Preserving Wireless Federated Learning Exploiting Inherent Hardware Impairments}

\author{\IEEEauthorblockN{Sina Rezaei Aghdam$^\text{1}$, Ehsan Amid$^\text{2}$, Marija Furdek$^\text{1}$, and~ Alexandre Graell i Amat$^\text{1}$} \\[-0.3cm]
\thanks{The research in this paper has been supported by Chalmers Artificial Intelligence Research Centre (CHAIR).
}
\IEEEauthorblockA{
\small $^\text{1}$\textit{Chalmers University of Technology, Gothenburg, Sweden}\\$^\text{2}$\textit{Google Research, Mountain View, CA}
}
}

\maketitle

\begin{abstract}
We consider a wireless federated learning system where multiple data holder edge devices collaborate to train a global model via sharing their parameter updates with an honest-but-curious parameter server.
We demonstrate that the inherent hardware-induced distortion perturbing the model updates of the edge devices can be exploited as a privacy-preserving mechanism.
In particular, we model the distortion as power-dependent additive Gaussian noise and present a power allocation strategy that provides privacy guarantees within the framework of differential privacy.
We conduct numerical experiments to evaluate the performance of the proposed power allocation scheme under different levels of hardware impairments.

\end{abstract}


\IEEEpeerreviewmaketitle

\section{Introduction}\label{sec:Intro}
\vspace{0.5em}

Federated learning has recently emerged as a promising technique for distributed machine learning in a variety of applications, including mobile edge computing \cite{lim20}.
Unlike conventional centralized machine learning techniques that require datasets to reside on a single server, federated learning leverages the local processing capabilities of edge devices and requires only a limited amount of communication between the devices and the server.
In particular, instead of offloading the entire data, each user uses its own dataset to train a local model and sends a small update to the global model maintained by the server.
This offers significant gains in terms of privacy and communication efficiency \cite{kairouz19}.

When deployed over wireless networks, the superposition nature of wireless channels can be used as a natural data aggregator for federated learning.
This phenomenon is referred to as \emph{over-the-air aggregation}, where simultaneously transmitted analog waves from different edge devices are weighted by channel coefficients and superposed at the server\cite{yang2020federated}.
Accurate and fair computation of the global model requires that the gradient estimates received from different devices have the same power at the parameter server.
One way to address this issue is to perform channel-inversion power control at the edge devices \cite{zhu2019broadband, amiri2020TWC}.
Alternatively, the parameter server can mitigate channel fading by using multiple receiver antennas \cite{Amiri_blind, Tegin2020}.

Despite its various attractive features, wireless federated learning in its primary form does not provide sufficiently strong privacy guarantees.
Although edge devices do not explicitly share their data in its original format, several recent works have shown that model parameters can themselves be informative to an honest-but-curious server (see, e.g., \cite{fredrikson2015, melis19, yin2021see, geiping2020inverting, mo2021quantifying}), leading to privacy leakage.
Accordingly, the study of privacy for wireless federated learning has been the subject of much recent research \cite{Seif20, Wei20, koda20, Liu20}. 
These works adopt the notion of differential privacy as a formal model for quantifying information disclosure about individuals and introduce strategies for limiting it.
Specifically, in \cite{Seif20, Wei20}, the authors show that by adding artificial noise to the model parameters on the edge device and properly adjusting the variance of this noise, different levels of differential privacy  protection can be achieved.
Moreover, \cite{koda20, Liu20} show that the inherent channel noise can be harnessed via proper power allocation to achieve privacy for free.

In this paper, we demonstrate that in addition to channel noise, the distortion introduced by the imperfect hardware of edge devices can also be used as a privacy-preserving mechanism.
Motivated by the analytical analysis and experimental validations in \cite{bjornson14a, schenk2008rf, studer10b}, we model distortion as an additive Gaussian noise whose variance is proportional to the signal power.
Within the differential privacy framework, we derive an upper-bound on the privacy violation probability and use it for formulating a privacy preservation condition.
We then propose a distortion-aware power allocation scheme to satisfy this condition.
We conduct numerical experiments to evaluate the performance of the proposed scheme under different levels of hardware impairments.
Our numerical results reveal that in realistic scenarios with imperfect hardware, taking into account the impact of distortion in power allocation can considerably reduce the performance loss caused by enforcing differential privacy constraints.
Exploiting distortion in realistic hardware can even result in improved performance compared to ideal distortion-free hardware.

The remainder of this paper is organized as follows.
In Section \ref{sec:SM}, we introduce our system model and the required technical background, including the definition of differential privacy.
The derivation of the privacy preservation criterion is described in Section \ref{sec:DP}.
In Section \ref{sec:PA}, we present a power allocation scheme that satisfies the differential privacy criterion.
In Section \ref{sec:Num}, we evaluate the performance of the proposed solution using numerical examples.
Finally, we conclude the paper in Section \ref{sec:Conc}.

\textit{Notation:} Throughout the paper, $\mathbb{R}$ and $\mathbb{C}$ represent the sets of real and complex values, respectively. We denote a zero-mean normal distribution with variance $\sigma^2$ by $\mathcal{N}(0,\sigma^2)$. Zero-mean complex Gaussian random variables are represented by $\mathcal{CN}(0,\sigma^2)$. The cardinality of a set $\mathcal{A}$ is denoted by $|\mathcal{A}|$. Furthermore, the difference between two sets is defined as $\mathcal{A}'-\mathcal{A}'' = \{x~|~x \in \mathcal{A}', x \not\in \mathcal{A}'' \}$ and $\|\mathcal{A}'-\mathcal{A}''\|_{1}$ denotes its cardinality. The union of sets $\mathcal{A}_1,\dots,\mathcal{A}_K$ is represented by $\mathcal{A} = \cup_{k=1}^{K}\mathcal{A}_k$. Vectors are denoted by boldfaced letters. The transpose and the Euclidean norm of a vector $\boldsymbol{a}$ are denoted by $\boldsymbol{a}^{\mathsf{T}}$ and $\|\boldsymbol{a}\|$, respectively.

\section{System Model}\label{sec:SM}

\begin{figure}[t] 
	\centering
	\noindent
	\includegraphics[width=1\columnwidth]{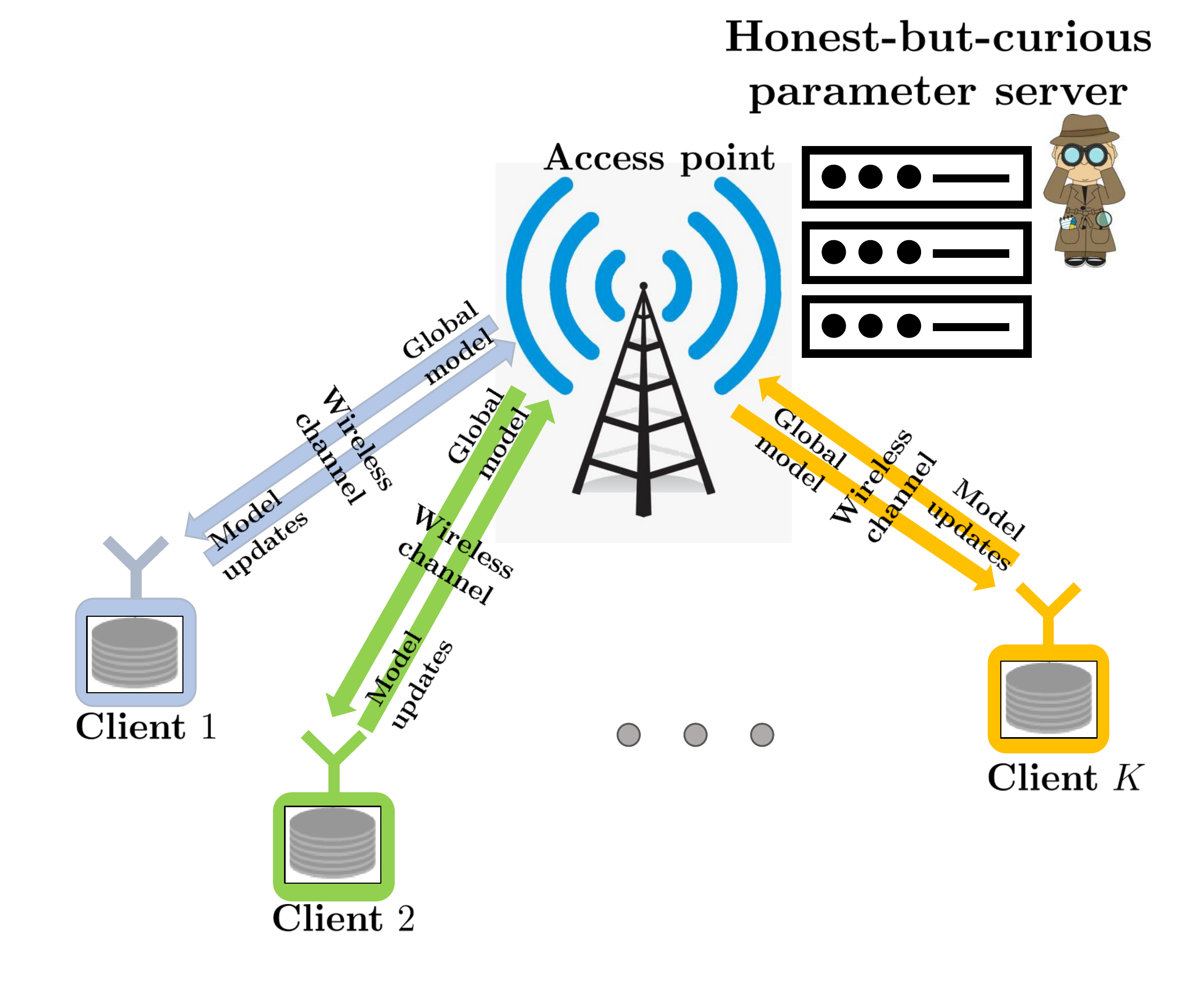}
	\caption{Wireless federated learning with $K$ edge devices and an honest-but-curious parameter server.}
	\label{system_model}
\end{figure}

We consider a wireless federated learning system (as shown in Fig. \ref{system_model}) consisting of $K$ single-antenna edge devices that attempt to collaboratively learn a shared model with the aid of a single-antenna parameter server.
Each device $k$ stores a local dataset, denoted by $\mathcal{B}_k$, which consists of
\begin{equation}\label{eq:dataset}
\mathcal{B}_k = \Big\{\big(\boldsymbol{u}_{k, i}, \upsilon_{k,i}\big)\Big\}_{i=1}^{|\mathcal{B}_k|},
\end{equation}
where $\boldsymbol{u}_{k, i}$ denotes the $i$-th training sample at the $k$-th device and $\upsilon_{k,i} \in \mathbb{R}$ is its corresponding label.
The federated learning process finds the optimal model parameters $\boldsymbol{\theta} \in \mathbb{R}^{d}$ that minimize the global loss function $F(\boldsymbol{\theta})$:
\begin{equation}\label{eq:min_global}
\underset{\boldsymbol{\theta}}{\text{minimize}}~~ F(\boldsymbol{\theta}) \overset{\Delta}{=} \frac{1}{K} \sum_{k = 1}^{K} F_k(\boldsymbol{\theta}).
\end{equation}
Here, $F_k(\boldsymbol{\theta})$ is the local loss function given by
\begin{equation}
    F_k(\boldsymbol{\theta}) = \frac{1}{|\mathcal{B}_k|} \sum_{(\boldsymbol{u}_{k, i}, \upsilon_{k,i}) \in \mathcal{B}_k} f(\boldsymbol{\theta}, \boldsymbol{u}_{k, i}, \upsilon_{k,i}),
\end{equation}
and $f(\boldsymbol{\theta}, \boldsymbol{u}_{k, i}, \upsilon_{k,i})$ is the sample loss that quantifies the prediction error of the model $\boldsymbol{\theta}$ on the training samples $\boldsymbol{u}_{k, i}$ with respect to the labels $\upsilon_{k,i}$.

\subsection{Learning Protocol}
In order to solve \eqref{eq:min_global}, a distributed stochastic gradient descent (SGD) is adopted by the parameter server and the edge devices via the following procedure:
\begin{itemize}
    \item \textbf{Step 1:} The parameter server broadcasts an initial model, i.e., $\boldsymbol{\theta}^{(1)}$, to the edge devices allowing them to initialize their learning model.
    \item \textbf{Step 2:} Each device performs local training on its own dataset and transmits the trained learning model parameters to the server. More specifically, at each communication round $t = 1, \dots, T$, each device computes the local gradient
    \begin{equation}\label{eq:local_grad}
           \nabla F_k(\boldsymbol{\theta}^{(t)})  =  \frac{1}{|\mathcal{B}_k|} \sum_{(\boldsymbol{u}_{k, i}, \upsilon_{k,i}) \in \mathcal{B}_k}  \nabla f(\boldsymbol{\theta}^{(t)}, \boldsymbol{u}_{k, i}, \upsilon_{k,i}),
    \end{equation}
    and sends it to the server. Here, $F_k(\boldsymbol{\theta}^{(t)})$ represents the local loss function.
    \item \textbf{Step 3:} The transmitted updates are aggregated over the air and received by the server. In particular the received aggregated signal at the parameter server, i.e., $\widehat{\nabla F}(\boldsymbol{\theta}^{(t)})$, yields an estimation of the global gradient. This allows the server to update the global model using gradient descent
    \begin{equation}\label{eq:model_update}
        \boldsymbol{\theta}^{(t+1)} = \boldsymbol{\theta}^{(t)} - \eta \widehat{\nabla F}(\boldsymbol{\theta}^{(t)}),
    \end{equation}
    where $\eta$ denotes the learning rate. The updated model is then broadcasted back to the edge devices.
    \item \textbf{Step 4:} Steps 2 and 3 are repeated until a convergence criterion is met.
\end{itemize}

\subsection{Communication Model}
All edge devices communicate with the edge server over a shared wireless channel. We assume an ideal downlink channel where the server broadcasts the initial and the aggregated models (Steps 1 and 3) through a distortion-free channel.
This assumption can be justified by the fact that the edge server can communicate through a base station, which usually has less stringent constraints in terms of power and hardware complexity compared to edge devices.
In the uplink transmission (Step 2), in order to facilitate the over-the-air aggregation of model updates, the local gradients are amplitude-modulated at the devices and sent to the server using an analog transmission scheme \cite{amiri2020SP}.
The non-ideal hardware in the edge devices distorts the transmit signal. 
We model this distortion as power-dependent additive Gaussian noise.
At communication round $t = 1, \dots, T$, the transmitted signal from the $k$-th device can be written as
\begin{equation}\label{eq:x_t}
    \boldsymbol{x}_k^{(t)} = \sqrt{\frac{\rho_k^{(t)}}{\| \boldsymbol{g}_k^{(t)} \|^2}} \boldsymbol{g}_k^{(t)} + \boldsymbol{e}_k^{(t)},
\end{equation}
where $\rho_k^{(t)}$ denotes the transmit power of device $k$ at communication round $t$, and $\boldsymbol{g}_k^{(t)} = |\mathcal{B}_k| \nabla F_k\big(\boldsymbol{\theta}^{(t)}\big)$ is the scaled version of the local gradient in \eqref{eq:local_grad}. 
The additive distortion noise $\boldsymbol{e}_k^{(t)}$ is distributed according to $\mathcal{N}\big(\textbf{0}, \kappa_k \rho_k^{(t)} \textbf{I}\big)$\footnote{Gaussianity can be justified analytically by the central limit theorem, since $\boldsymbol{e}_k^{(t)}$ describes the aggregate effect of different residual hardware impairments \cite{studer10b}.} where $\kappa_k \in [0, 1)$ is the proportionality coefficient.
The rationale behind this model is that, in many practical cases, a fixed portion of the signal is turned into distortion \cite{bjornson14a}. 
This can happen, for example, due to amplitude-to-amplitude (AM-AM) nonlinearities in the power amplifier \cite{schenk2008rf}.

We further assume that the edge devices have perfect channel state information and accordingly, they compensate for the phases of their channels in the course of transmission. The received signal at the server can therefore be written as
\begin{equation}\label{eq:y_t}
    \boldsymbol{y}^{(t)} = \sum_{k = 1}^{K} \big|h_k^{(t)}\big| \boldsymbol{x}_k^{(t)} + \boldsymbol{w}^{(t)},
\end{equation}
where $h_k^{(t)}$ denotes the fading channel coefficient for the $k$-th device at the $t$-th communication round, and $\boldsymbol{w}^{(t)} \sim \mathcal{N}(0, N_0)$ is independent and identically distributed (i.i.d.) additive Gaussian noise.
We consider the case of block fading channels where the channel coefficients remain unchanged within each time slot, and change independently from one communication round to the other.

\subsection{Threat Model and Privacy Mechanism}
We consider an honest-but-curious server that follows the protocol instructions correctly, but it may attempt to break privacy through observing the signals received throughout the learning process.
Specifically, the server observes
\begin{equation}\label{eq:y_t_2}
     \boldsymbol{y}^{(t)} = \sum_{k = 1}^{K} b_k^{(t)} \big|h_k^{(t)}\big| \boldsymbol{g}_k^{(t)} + \boldsymbol{w}_{\text{eff}}^{(t)},
\end{equation}
over $t = 1, \dots, T$, and can infer information about the edge devices' private data.
Here, $b_k^{(t)}$ is given by
\begin{equation}\label{eq:b_k^t}
   b_k^{(t)} = \sqrt{\frac{\rho_k^{(t)}}{\| \boldsymbol{g}_k^{(t)} \|^2}},
\end{equation}
and $\boldsymbol{w}_{\text{eff}}^{(t)} = \boldsymbol{w}^{(t)} + \sum_{k = 1}^{K} \big|h_k^{(t)}\big| \boldsymbol{e}_k^{(t)}$ is the effective noise distributed according to $\mathcal{N}\big(0, (\sigma^{(t)})^2\big)$, with
\begin{equation}\label{eq:std}
\sigma^{(t)} = \sqrt{N_0 + \sum_{k=1}^{K} \kappa_k \rho_k^{(t)} |h_k^{(t)}|^2}.
\end{equation}
The effective noise $\boldsymbol{w}_{\text{eff}}^{(t)}$ will be exploited as a privacy-preserving mechanism in the next section.

\subsection{Differential Privacy} 
Differential privacy is a strong and provable privacy guarantee for an individual's input to a randomized function, i.e., a privacy mechanism.
Informally, this guarantee implies that the behavior of the mechanism is essentially unchanged over inputs from any pair of adjacent datasets.
In our case, adjacent datasets are the same, except for an example associated with a single edge device.
Mathematically, this can be expressed as two global datasets $\mathcal{B}' = \cup_{k=1}^{K} \mathcal{B}_k'$ and $\mathcal{B}'' = \cup_{k=1}^{K} \mathcal{B}_k''$ where $\|\mathcal{B}_i'-\mathcal{B}_i''\|_1 = 1$ for some device $i$ and   $\|\mathcal{B}_k'-\mathcal{B}_k''\|_1 = 0$ for all $k = 1, \dots, K$ except for $k = i$ \cite{Liu20}.
Using this definition, we can now formally define the notion of $(\epsilon, \delta)$ differential privacy.
\begin{mydef}[Differential Privacy \cite{DP_book14, Liu20}]
Let $\epsilon > 0 $ and $0 \leq \delta \leq 1$. For any two adjacent datasets $\mathcal{B}'$ and $\mathcal{B}''$, $(\epsilon, \delta)$ differential privacy requires that
\begin{equation} \label{eq:DP}
			\text{P}(\boldsymbol{y}|\mathcal{B}') \leq \exp{(\epsilon)} \text{P}(\boldsymbol{y}|\mathcal{B}'') + \delta,
\end{equation}
where $\text{P}(\boldsymbol{y}|\mathcal{B}')$ and $\text{P}(\boldsymbol{y}|\mathcal{B}'')$ stand for the distributions of received signals $\boldsymbol{y} = \big\{\boldsymbol{y}^{(t)}\big\}_{t = 1}^{T}$ conditioned on the uses of either of the adjacent datasets.
\end{mydef}

The $(\epsilon, \delta)$ differential privacy condition in \eqref{eq:DP} can be rewritten as \cite[Lemma 3.17]{DP_book14} 
\begin{equation}\label{eq:DP_loss_prob}
    \text{Pr}(| \mathcal{L}_{\mathcal{B}',\mathcal{B}''}(\boldsymbol{y}) | \leq \epsilon) \geq 1-\delta,
\end{equation}
where
\begin{equation}\label{eq:DPLoss}
    	\mathcal{L}_{\mathcal{B}',\mathcal{B}''}(\boldsymbol{y}) = \ln \bigg( \frac{	\text{P}(\boldsymbol{y}|\mathcal{B}')}{\text{P}(\boldsymbol{y}|\mathcal{B}'')}\bigg)
\end{equation}
is referred to as differential privacy loss.
According to \eqref{eq:DP_loss_prob}, $(\epsilon, \delta)$ differential privacy ensures that the absolute value of the privacy loss is bounded by $\epsilon$ with probability at least $1-\delta$.
Smaller values of $\epsilon$ and $\delta$ provide therefore stronger privacy guarantees.
More precisely, if \eqref{eq:DP_loss_prob} is satisfied for sufficiently small $\epsilon$ and $\delta$, this implies that it is statistically impossible for the parameter server to detect whether a particular training sample is included in the training dataset, even in the extreme case when all other training samples are known.

\section{Privacy Preservation Condition}\label{sec:DP}
In this section, we derive a privacy preservation condition by applying the notion of differential privacy to our problem.
To this end, we first derive an expression for the privacy loss due to the disclosure of the received signal $\boldsymbol{y} = \big[\boldsymbol{y}^{(1)}, \dots, \boldsymbol{y}^{(T)} \big]$ at the parameter server by exploiting inherent distortion as the privacy-preserving mechanism.
This can be written as
\begin{align} \label{eq:PLoss_2}
\nonumber
	\mathcal{L}_{\mathcal{B}',\mathcal{B}''}(\boldsymbol{y}) &= \ln \Bigg(\prod_{t = 1}^{T}  \frac{\text{P}(\boldsymbol{y}^{(t)}|\boldsymbol{y}^{(t-1)},\dots, \boldsymbol{y}^{(1)},\mathcal{B}')}{\text{P}(\boldsymbol{y}^{(t)}|\boldsymbol{y}^{(t-1)},\dots, \boldsymbol{y}^{(1)},\mathcal{B}'')} \Bigg)\\ 
	&= \sum_{t=1}^{T} \ln \Bigg( \frac{\text{P}(\boldsymbol{y}^{(t)}|\boldsymbol{y}^{(t-1)},\dots, \boldsymbol{y}^{(1)},\mathcal{B}')}{\text{P}(\boldsymbol{y}^{(t)}|\boldsymbol{y}^{(t-1)},\dots, \boldsymbol{y}^{(1)},\mathcal{B}'')} \Bigg).
\end{align}
From \eqref{eq:y_t_2} and for a given dataset $\mathcal{B} = \cup_{k=1}^{K} \mathcal{B}_k$, we have
\begin{align} \label{eq:ConditionalPDF} \nonumber
    \text{P}&\big(\boldsymbol{y}^{(t)}|\boldsymbol{y}^{(t-1)},\dots,  \boldsymbol{y}^{(1)},\mathcal{B}\big) =\frac{1}{\sigma^{(t)} \sqrt{2 \pi}} \\ & \cdot \exp  \left(-\frac{\Big\|\boldsymbol{y}^{(t)} - \sum_{k = 1}^{K} b_k^{(t)}(\mathcal{B}_k) \big|h_k^{(t)}\big|  \boldsymbol{g}_k^{(t)}(\mathcal{B}_k)\Big\|^2}{2 (\sigma^{(t)})^2 }\right),
\end{align}
where $b_k^{(t)}(\mathcal{B}_k)$ can be calculated by replacing $\boldsymbol{g}_k^{(t)}(\mathcal{B}_k)$ in \eqref{eq:b_k^t}.
Now, by using \eqref{eq:ConditionalPDF} in \eqref{eq:PLoss_2} we obtain
\begin{align}\label{eq:PLoss_3}
    \nonumber
	& \mathcal{L}_{\mathcal{B}',\mathcal{B}''}(\boldsymbol{y})\\ 
	\nonumber &~~= \sum_{t=1}^{T} \ln \left( \frac{ \exp\left(- \frac{\Big\|\boldsymbol{y}^{(t)} - \sum_{k = 1}^{K} b_k^{(t)}(\mathcal{B}_k') \left|h_k^{(t)}\right|  \boldsymbol{g}_k^{(t)}(\mathcal{B}_k')\Big\|^2}{2 (\sigma^{(t)})^2}   \right)}{\exp\left(- \frac{\Big\|\boldsymbol{y}^{(t)} - \sum_{k = 1}^{K} b_k^{(t)}(\mathcal{B}_k'') \left|h_k^{(t)}\right|  \boldsymbol{g}_k^{(t)}(\mathcal{B}_k'')\Big\|^2}{2 (\sigma^{(t)})^2}   \right)} \right) \\ \nonumber
	&~~= \sum_{t = 1}^{T} \ln \left( \frac{ \exp\left(- \frac{\|\boldsymbol{w}_{\text{eff}}^{(t)}\|^2}{2 (\sigma^{(t)})^2}   \right)}{\exp\left(- \frac{\|\boldsymbol{w}_{\text{eff}}^{(t)} + \boldsymbol{v}^{(t)} \|^2}{2 (\sigma^{(t)})^2}   \right)} \right)\\ 
	&~~= \sum_{t=1}^{T} \frac{\|\boldsymbol{v}^{(t)}\|^2 + 2 (\boldsymbol{w}_{\text{eff}}^{(t)})^{\mathsf{T}}\boldsymbol{v}^{(t)} }{2 (\sigma^{(t)})^2} \overset{\Delta}{=} \Gamma, 
\end{align}
where
\begin{equation}\label{vt}
    \boldsymbol{v}^{(t)} = \sum_{k = 1}^{K}  \big|h_k^{(t)}\big| \Big(b_k^{(t)}(\mathcal{B}_k'') \boldsymbol{g}_k^{(t)}(\mathcal{B}_k'') - b_k^{(t)}(\mathcal{B}_k') \boldsymbol{g}_k^{(t)}(\mathcal{B}_k')\Big).
\end{equation}
By applying the triangle inequality, the norm of $\boldsymbol{v}^{(t)}$ can be bounded as
\begin{equation}\label{eq:v_norm}
    \|\boldsymbol{v}^{(t)}\| \leq 2 \max_{k} \sqrt{\rho_k^{(t)}} \big|h_k^{(t)}\big|   \overset{\Delta}{=} \Delta^{(t)}.
\end{equation}
Following similar steps as in \cite[Appendix A]{DP_book14} and \cite[Eq. (52)]{Liu20}, the following upper-bound is obtained on the privacy violation probability
\begin{align}\label{eq:violprb} \nonumber
      \text{Pr}\left(   \left| \Gamma \right| > \epsilon  \right) &\overset{(\text{i})}{\leq} \text{Pr}\left( \left|\sum_{t=1}^{T}  \frac{ (\boldsymbol{w}_{\text{eff}}^{(t)})^{\mathsf{T}}\boldsymbol{v}^{(t)} }{ (\sigma^{(t)})^2} \right| > \epsilon - \sum_{t=1}^{T} \frac{\|\boldsymbol{v}^{(t)}\|^2}{2 (\sigma^{(t)})^2}  \right)\\ \nonumber
    &= 2 \text{Pr}\left( \sum_{t=1}^{T}  \frac{ (\boldsymbol{w}_{\text{eff}}^{(t)})^{\mathsf{T}}\boldsymbol{v}^{(t)} }{ (\sigma^{(t)})^2}  > \epsilon - \sum_{t=1}^{T} \frac{\|\boldsymbol{v}^{(t)}\|^2}{2 (\sigma^{(t)})^2}  \right)\\ 
    &\overset{(\text{ii})}{\leq} 2 \text{Pr}\left( \Lambda > \epsilon - \sum_{t = 1}^{T} \frac{1}{2} \left( \frac{\Delta^{(t)}}{\sigma^{(t)}} \right)^2 \right),
\end{align}
where (i) comes from the inequality $\text{Pr}(|X+c|> \epsilon) \leq \text{Pr}(|X|+c> \epsilon)$ for any arbitrary $c \geq 0$, and (ii) is obtained using \eqref{eq:v_norm} where
\begin{equation}
 \Lambda \sim \mathcal{CN}\left(0,\sum_{t = 1}^{T} \left( \frac{\Delta^{(t)}}{\sigma^{(t)}} \right)^2\right).
\end{equation}
%
The upper-bound in \eqref{eq:violprb} can therefore be written as
\begin{align}\label{eq:bound_expression}\nonumber
 &\text{Pr}\left( \left| \Gamma \right| > \epsilon  \right)\\ ~~~&\leq
 \frac{2}{\sqrt{2 \pi \sum_{t = 1}^{T} \left( \frac{\Delta^{(t)}}{\sigma^{(t)}} \right)^2 }} \hspace{-0.2em} \int_s^{\infty} \hspace{-0.8em} \exp \left(-\frac{z^2}{2 \sum_{t = 1}^{T} \left( \frac{\Delta^{(t)}}{\sigma^{(t)}} \right)^2 }\right) \text{d}z,
\end{align}
where
\begin{equation}\label{eq:s}
    s = \epsilon - \sum_{t = 1}^{T} \frac{1}{2} \left( \frac{\Delta^{(t)}}{\sigma^{(t)}} \right)^2 > 0. 
\end{equation}
Finally, by using \eqref{eq:violprb}-\eqref{eq:s} in \eqref{eq:DP_loss_prob}, the privacy preservation condition is expressed as
\begin{equation}\label{eq:DPcondition}
     \frac{2}{\sqrt{2 \pi \nu }} \int_{\epsilon - \frac{\nu}{2}}^{\infty} \exp \left(-\frac{z^2}{2 \nu}\right) \text{d}z < \delta,
\end{equation}
where
\begin{equation}\label{eq:nu}
    \nu = \sum_{t=1}^{T} \left( \frac{\Delta^{(t)}}{\sigma^{(t)}} \right)^2 = \sum_{t=1}^{T} \frac{\left(2 \max_{k} \sqrt{\rho_k^{(t)}} \big|h_k^{(t)}\big| \right)^2}{N_0 + \sum_{k=1}^{K} \kappa_k \rho_k^{(t)} |h_k^{(t)}|^2}.
\end{equation}
\vspace{0.7em}

\section{Privacy-Preserving Power Allocation Scheme}\label{sec:PA}

In this section, we present an offline-optimized privacy-preserving power allocation scheme.
In particular, we assume that the parameters $\{h_k^{(t)}, \kappa_k, N_0 \}$ are known and we seek the values $\rho_k^{(t)}$ for $k=1, \dots, K$ and $t = 1, \dots, T$ to minimize the distortion of the recovered gradient from the received signal in \eqref{eq:y_t_2} with respect to the ground-truth global gradient, i.e., 
\begin{align}\label{eq:MSE}
    & \text{MSE}^{(t)} = \mathbb{E} \left[ \left\| \widehat{\nabla F}(\boldsymbol{\theta}^{(t)}) - {\nabla F}(\boldsymbol{\theta}^{(t)})  \right\|^2 \right] \nonumber \\
    & = \frac{1}{K^2} \mathbb{E} \left[ \left\| \sum_{k = 1}^{K} \frac{\sqrt{\rho_k^{(t)}} \big|h_k^{(t)}\big| \boldsymbol{g}_k^{(t)}}{\xi^{(t)} \big\| \boldsymbol{g}_k^{(t)} \big\|} + \frac{\boldsymbol{w}_{\text{eff}}^{(t)}}{\xi^{(t)}} - \sum_{k = 1}^{K}  \frac{\boldsymbol{g}_k^{(t)}}{\big\| \boldsymbol{g}_k^{(t)} \big\|}   \right\|^2 \right],
\end{align}
while satisfying the differential privacy condition in \eqref{eq:DPcondition}.
In \eqref{eq:MSE}, the expectation is over the distribution of the effective noise $\boldsymbol{w}_{\text{eff}}$ and $\xi^{(t)}$ is the normalization factor applied by the parameter server for recovering the gradient of interest. 
We assume that the edge devices are subject to a peak power constraint, i.e.,
\begin{equation}\label{eq:MaxPower}
    (1 + \kappa_k) \rho_k^{(t)} \leq \rho_{\text{max}}.
\end{equation}
Accordingly, the relevant optimization problem can be defined as
\begin{IEEEeqnarray}{lCr} \label{eq:Optimization}
\underset{\rho_k^{(t)}}{\text{minimize}}  ~~&\text{MSE}^{(t)}\\
\text{subject to}~~  &\eqref{eq:DPcondition}~\text{and}~ \eqref{eq:MaxPower}. \nonumber
\end{IEEEeqnarray}

To guarantee a perfectly unbiased global model, the power scaling factors should be set such that the following gradient alignment condition is satisfied,
\begin{equation}\label{eq:no_bias}
    \sqrt{\rho_k^{(t)}} \big|h_k^{(t)}\big|  = \lambda^{(t)}.
\end{equation}
In this case, by setting the normalization factor $\xi^{(t)} = \lambda^{(t)}$, the mean squared error in \eqref{eq:MSE} simplifies to
\begin{equation}\label{eq:MSE_nobias}
    \text{MSE}^{(t)} = \mathbb{E} \left[ \left\|  \frac{\boldsymbol{w}_{\text{eff}}^{(t)}}{\lambda^{(t)}} \right\|^2 \right] = \frac{N_0 + \sum_{k=1}^{K} \kappa_k \lambda^{(t)} }{ K^2 (\lambda^{(t)})^2}.
\end{equation}

When no privacy constraint is in place, the optimal solution for minimizing \eqref{eq:MSE_nobias} under the power constraint in \eqref{eq:MaxPower} is given by \cite{Xcao2020}
\begin{equation}\label{eq:Opt_Pow_noPriv}
    \tsup[1]{\rho}_{k}^{(t)} =  \frac{\rho_{\text{max}} \big|h_j^{(t)}\big|^2 }{(1 + \kappa_{j}) \big|h_k^{(t)}\big|^2},
\end{equation}
where $j$ is the index of the weakest channel, i.e.,
\begin{equation}
    j = \argmin_{k} \big|h_k^{(t)}\big|.
\end{equation}
In words, the user with the weakest channel performs a full power transmission while the others apply power control with channel inversion to satisfy \eqref{eq:no_bias} using the resulting $\tsup[1]{\lambda}^{(t)} = \sqrt{\rho_{\text{max}} \big|h_j^{(t)}\big|^2 / (1 + \kappa_{j})}$. 
Note that when differential privacy is enforced, due to the constraint \eqref{eq:DPcondition},  $\tsup[1]{\lambda}^{(t)}$ may no longer be a feasible solution.
In this case, noting that the privacy violation probability is an increasing function of $\lambda^{(t)}$, the value of $\tsup[1]{\lambda}^{(t)}$ should be decreased until the constraint \eqref{eq:DPcondition} is satisfied.
Denote the maximum values of $\lambda^{(t)}$ satisfying \eqref{eq:DPcondition} by $\lambda_{\text{p}}^{(t)}$.
The solution to the optimization problem in \eqref{eq:Optimization} can then be stated as
\begin{equation}\label{eq:Optim_Pow}
    \breve{\rho}_{k}^{(t)} =  \frac{ \big|\breve{\lambda}^{(t)}\big|^2 }{ \big|h_k^{(t)}\big|^2},
\end{equation}
where
\begin{equation}
    \breve{\lambda}^{(t)} = \min \left\{\tsup[1]{\lambda}^{(t)}, \lambda_{\text{p}}^{(t)} \right\},
\end{equation}
where $\lambda_{\text{p}}^{(t)}$ can be obtained using a simple line search.

\section{Numerical Results}\label{sec:Num}

In this section, we evaluate the performance of the proposed privacy-preserving power allocation schemes under different hardware impairment levels.\footnote{For the sake of reproducibility, the codes used to obtain the simulation results will be made available after the review process is completed.}
We consider the learning task of handwritten digit recognition using the well-known MNIST dataset, which consists of $60000$ training samples and $10000$ test samples. 
Each example is a $28 \times 28$ pixel gray scale image.
The training samples are evenly distributed among the $K$ devices.
As a baseline classification model, we consider a feedforward neural network with a single hidden layer containing $100$ hidden units followed by a softmax output layer.
We use Adam as the optimizer with an initial learning rate of $\eta = 0.001$; the optimizer automatically adjusts the learning rate as the number of learning iterations increases to achieve faster convergence.
We assume that the number of local steps an edge device performs before transmitting an update is equal to $30$.
The batch size is set to $128$.

\begin{figure}[t!] 
	\centering
	\noindent
	\includegraphics[width=9cm]{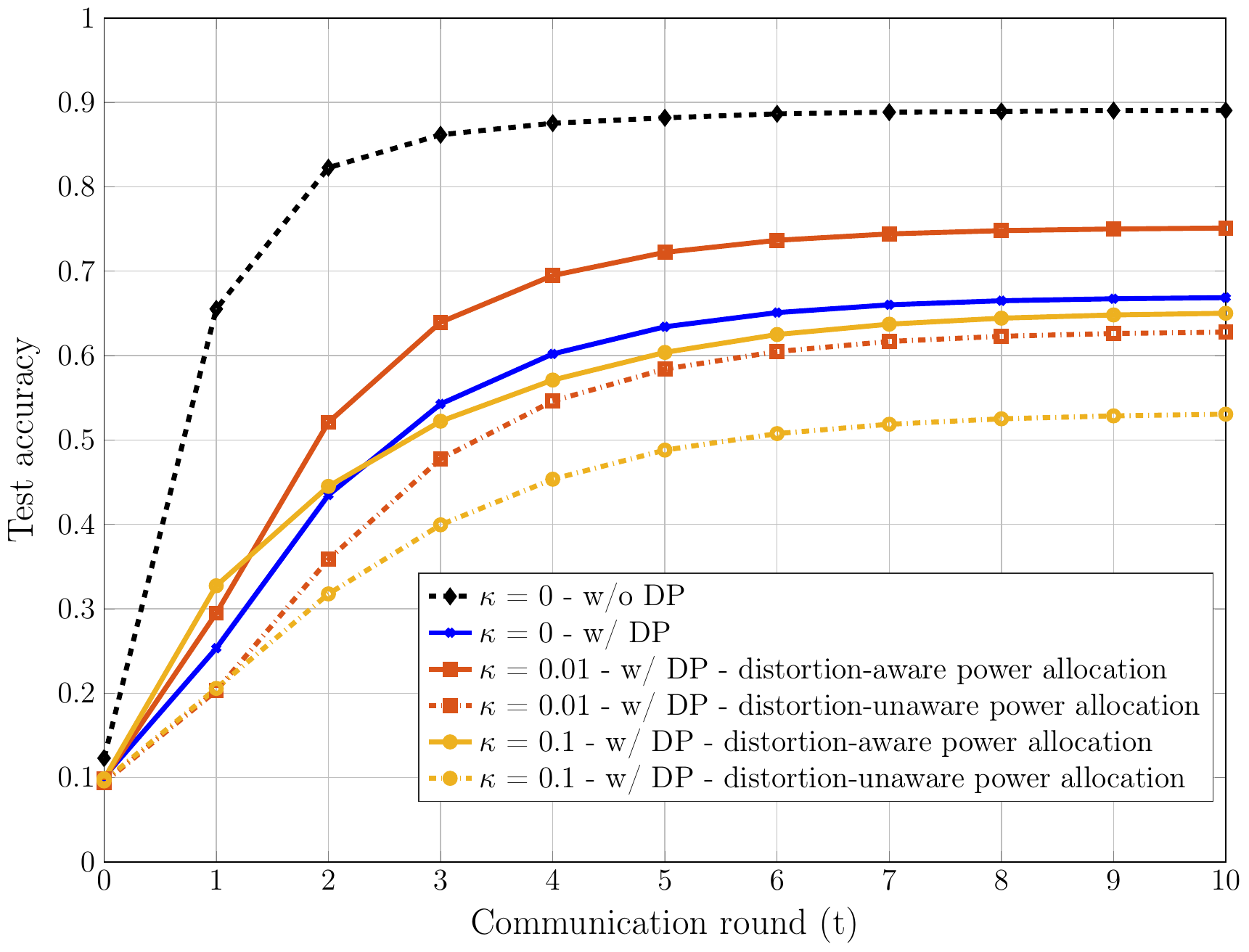}
	\vspace{0.3em}
	\caption{Test accuracy for different power allocation strategies under several hardware impairment levels.}
	\label{fig_Test_Acc}
\end{figure}

We consider a scenario with $K = 50$ edge devices and set the number of communication rounds to $T=10$.
The channel coefficients $h_k^{(t)}$ are drawn from a complex circularly-symmetric Gaussian distribution.
The peak power constraint is set to $\rho_{\text{max}} = 10$\,dBm for all devices. The receiver noise variance is $N_0 = -20$\,dBm and the differential privacy parameters are set to $\epsilon = 25$ and $\delta = 0.05$.
Finally, we assume that the proportionality coefficient is equal for all edge devices, i.e., $\kappa_k = \kappa$ for $k = 1,\dots,K$. We consider three different hardware impairment levels, $\kappa  \in \{0, 0.01, 0.1\}$.
Note that $\kappa = 0$ corresponds to a scenario with ideal hardware, whereas $\kappa = 0.01$ and $\kappa = 0.1$ represent cases with low-to-moderate and high levels of hardware impairment, respectively.\footnote{The proportionality coefficient $\kappa$ is related to the transmit error vector magnitude (EVM) as $\text{EVM} = \sqrt{\kappa}$ \cite{bjornson14a}. Accordingly, $\kappa = 0.01$ and $\kappa = 0.1$ correspond to $10\%$ and $31.62\%$ EVM values, respectively. Typical EVM values can be derived from the minimum requirements specified in established standards.}
In each of these scenarios, we evaluate the learning performance versus the communication round.
As benchmarks, we also include the results for the cases with distortion-unaware power allocation (where the effect of hardware impairments is ignored and the power allocation is carried out considering $\kappa = 0$ regardless of its actual value) and the case with ideal hardware and no privacy constraint.

Fig. \ref{fig_Test_Acc} depicts the test accuracy, averaged over 50 independent trials with random splitting of the dataset in each trial, for the cases mentioned above.
From the figure, it can be seen that for ideal hardware, enforcing the privacy constraint leads to a significant degradation in learning performance. 
This is because in this case, receiver noise is the only privacy-preserving mechanism, and in order to satisfy the differential privacy constraint, devices must scale down their transmit power. 
However, by adopting the proposed power allocation schemes in realistic scenarios with imperfect hardware, the hardware-induced distortion at the devices can be used together with the receiver noise to satisfy the privacy requirements with a lower penalty in terms of learning performance.
The proposed power allocation strategy can, therefore, achieve considerable performance gains compared to the conventional strategies that ignore the effect of hardware-induced distortion.
In certain cases, the imperfect hardware devices can even achieve better performance than the ideal hardware scenario. 
An example of such gain attained by utilizing hardware-induced distortion can be seen for the case with $\kappa = 0.01$ (red curves with square markers).
In particular, with $\kappa = 0.01$, the devices can transmit more power and yet guarantee the privacy preserving condition in \eqref{eq:DPcondition}, resulting in improved learning performance despite the slightly increased effective noise variance.

It should be noted that whether the imperfect hardware yields an improved performance with respect to the ideal hardware depends on the level of impairments as well as the stringency of the privacy requirements.
For instance, in the example given in Fig. \ref{fig_Test_Acc}, it can be seen that the case with $\kappa = 0.1$ (yellow curves with circle markers) yields a degraded performance with respect to the ideal hardware.
The reason behind this is that the amount of distortion injected into the transmitted updates is more than enough for satisfying the differential privacy requirements.
As a result, the amount of transmit power at the devices is determined by the peak power constraint in \eqref{eq:MaxPower} and the excessive distortion degrades the learning performance.
Adopting a truncation-based approach by excluding the edge devices that experience deep fading channels similar to the solution in \cite{zhu2019broadband}, can reduce the amount of excessive distortion.
A detailed investigation of this approach is left for future research.

\section{Conclusion}\label{sec:Conc}

We studied differentially-private wireless federated learning in realistic scenarios where edge devices are equipped with imperfect hardware.
By modeling the hardware-induced distortion as power-dependent additive white Gaussian noise, we derived an expression for the privacy preservation condition and proposed a power allocation scheme to satisfy it.
Our numerical results indicate that exploiting the inherent distortion considerably reduces the performance loss caused by the enforcement of differential privacy constraints. 
As a result, our proposed power allocation scheme achieves significant gains in terms of learning performance compared to conventional approaches that ignore the effect of hardware-induced distortion.
%






%
\vspace{1em}
\begin{spacing}{1}
\bibliographystyle{IEEEtran}
\bibliography{IEEEabrv,confs-jrnls,publishers,wflbib}

\begin{thebibliography}{10}
\providecommand{\url}[1]{#1}
\csname url@samestyle\endcsname
\providecommand{\newblock}{\relax}
\providecommand{\bibinfo}[2]{#2}
\providecommand{\BIBentrySTDinterwordspacing}{\spaceskip=0pt\relax}
\providecommand{\BIBentryALTinterwordstretchfactor}{4}
\providecommand{\BIBentryALTinterwordspacing}{\spaceskip=\fontdimen2\font plus
\BIBentryALTinterwordstretchfactor\fontdimen3\font minus
  \fontdimen4\font\relax}
\providecommand{\BIBforeignlanguage}[2]{{%
\expandafter\ifx\csname l@#1\endcsname\relax
\typeout{** WARNING: IEEEtran.bst: No hyphenation pattern has been}%
\typeout{** loaded for the language `#1'. Using the pattern for}%
\typeout{** the default language instead.}%
\else
\language=\csname l@#1\endcsname
\fi
#2}}
\providecommand{\BIBdecl}{\relax}
\BIBdecl

\bibitem{lim20}
W.~Y.~B. Lim, N.~C. Luong, D.~T. Hoang, Y.~Jiao, Y.-C. Liang, Q.~Yang,
  D.~Niyato, and C.~Miao, ``Federated learning in mobile edge networks: A
  comprehensive survey,'' \emph{IEEE Commun. Surv. Tut.}, vol.~22, no.~3, pp.
  2031--2063, third quarter 2020.

\bibitem{kairouz19}
P.~Kairouz, H.~B. McMahan, B.~Avent, A.~Bellet, M.~Bennis, A.~N. Bhagoji,
  K.~Bonawitz, Z.~Charles, G.~Cormode, R.~Cummings \emph{et~al.}, ``Advances
  and open problems in federated learning,'' \emph{arXiv preprint
  arXiv:1912.04977}, 2019.

\bibitem{yang2020federated}
K.~Yang, T.~Jiang, Y.~Shi, and Z.~Ding, ``Federated learning via over-the-air
  computation,'' \emph{{IEEE} Trans. Wireless Commun.}, vol.~19, no.~3, pp.
  2022--2035, Jan. 2020.

\bibitem{zhu2019broadband}
G.~Zhu, Y.~Wang, and K.~Huang, ``Broadband analog aggregation for low-latency
  federated edge learning,'' \emph{{IEEE} Trans. Wireless Commun.}, vol.~19,
  no.~1, pp. 491--506, Jan. 2020.

\bibitem{amiri2020TWC}
M.~M. Amiri and D.~G{\"u}nd{\"u}z, ``Federated learning over wireless fading
  channels,'' \emph{{IEEE} Trans. Wireless Commun.}, vol.~19, no.~5, pp.
  3546--3557, May 2020.

\bibitem{Amiri_blind}
M.~M. Amiri, T.~M. Duman, D.~Gunduz, S.~R. Kulkarni, and H.~V. Poor, ``Blind
  federated edge learning,'' \emph{arXiv preprint arXiv:2010.10030}, 2020.

\bibitem{Tegin2020}
B.~Tegin and T.~M. Duman, ``Machine learning at wireless edge with ofdm and low
  resolution {ADC} and {DAC},'' \emph{arXiv preprint arXiv:2010.00350}, 2020.

\bibitem{fredrikson2015}
M.~Fredrikson, S.~Jha, and T.~Ristenpart, ``Model inversion attacks that
  exploit confidence information and basic countermeasures,'' in \emph{Proc.
  ACM SIGSAC Conf. Comput. Commun. Security}, 2015, pp. 1322--1333.

\bibitem{melis19}
L.~Melis, C.~Song, E.~De~Cristofaro, and V.~Shmatikov, ``Exploiting unintended
  feature leakage in collaborative learning,'' in \emph{Proc. IEEE Symp. Secur.
  Privacy (SP)}, May 2019, pp. 691--706.

\bibitem{yin2021see}
H.~Yin, A.~Mallya, A.~Vahdat, J.~M. Alvarez, J.~Kautz, and P.~Molchanov, ``See
  through gradients: Image batch recovery via gradinversion,'' in \emph{Proc.
  IEEE/CVF Conf. Computer Vision Pattern Recognition}, 2021, pp.
  16\,337--16\,346.

\bibitem{geiping2020inverting}
J.~Geiping, H.~Bauermeister, H.~Dr{\"o}ge, and M.~Moeller, ``Inverting
  gradients--how easy is it to break privacy in federated learning?''
  \emph{arXiv preprint arXiv:2003.14053}, 2020.

\bibitem{mo2021quantifying}
F.~Mo, A.~Borovykh, M.~Malekzadeh, H.~Haddadi, and S.~Demetriou, ``Quantifying
  information leakage from gradients,'' \emph{arXiv preprint arXiv:2105.13929},
  2021.

\bibitem{Seif20}
M.~Seif, R.~Tandon, and M.~Li, ``Wireless federated learning with local
  differential privacy,'' in \emph{Proc. IEEE Int. Symp. Inf. Theory (ISIT)},
  Los Angeles, CA, USA, Jun. 2020, pp. 2604--2609.

\bibitem{Wei20}
K.~Wei, J.~Li, M.~Ding, C.~Ma, H.~H. Yang, F.~Farokhi, S.~Jin, T.~Q. Quek, and
  H.~V. Poor, ``Federated learning with differential privacy: Algorithms and
  performance analysis,'' \emph{IEEE Trans. Inf. Forensics Security}, vol.~15,
  pp. 3454--3469, Apr. 2020.

\bibitem{koda20}
Y.~Koda, K.~Yamamoto, T.~Nishio, and M.~Morikura, ``Differentially private
  {AirComp} federated learning with power adaptation harnessing receiver
  noise,'' in \emph{Proc. IEEE Global Commun. Conf. (GLOBECOM)}, Taipei,
  Taiwan, Dec. 2020.

\bibitem{Liu20}
D.~{Liu} and O.~{Simeone}, ``Privacy for free: Wireless federated learning via
  uncoded transmission with adaptive power control,'' \emph{{IEEE} J. Sel.
  Areas Commun.}, vol.~39, no.~1, pp. 170--185, Jan. 2021.

\bibitem{bjornson14a}
E.~Bj{\"o}rnson, J.~Hoydis, M.~Kountouris, and M.~Debbah, ``Massive {MIMO}
  systems with non-ideal hardware: Energy efficiency, estimation, and capacity
  limits,'' \emph{{IEEE} Trans. Inf. Theory}, vol.~11, no.~60, pp. 7112--7139,
  Nov. 2014.

\bibitem{schenk2008rf}
T.~Schenk, \emph{RF imperfections in high-rate wireless systems: {I}mpact and
  digital compensation}.\hskip 1em plus 0.5em minus 0.4em\relax Springer
  Science \& Business Media, 2008.

\bibitem{studer10b}
C.~Studer, M.~Wenk, and A.~Burg, ``{MIMO} transmission with residual
  transmit-{RF} impairments,'' in \emph{Proc. Int. ITG Workshop on Smart
  Antennas (WSA)}, Bremen, Germany, Feb. 2010, pp. 189--196.

\bibitem{amiri2020SP}
M.~M. Amiri and D.~G{\"u}nd{\"u}z, ``Machine learning at the wireless edge:
  Distributed stochastic gradient descent over-the-air,'' \emph{IEEE Trans.
  Signal Process.}, vol.~68, pp. 2155--2169, Mar. 2020.

\bibitem{DP_book14}
C.~Dwork and A.~Roth, ``The algorithmic foundations of differential privacy,''
  \emph{Foundations and Trends in Theoretical Computer Science}, vol.~9, no.
  3-4, pp. 211--407, 2014.

\bibitem{Xcao2020}
X.~Cao, G.~Zhu, J.~Xu, and K.~Huang, ``Optimized power control for over-the-air
  computation in fading channels,'' \emph{{IEEE} Trans. Wireless Commun.},
  vol.~19, no.~11, pp. 7498--7513, Aug. 2020.

\end{thebibliography}
\end{spacing}

\end{document}